# Dynamical pinning of domain wall in magnetic nanowire induced by Walker breakdown

Hironobu Tanigawa, Tomohiro Koyama, Maciej Bartkowiak, Shinya Kasai, Kensuke Kobayashi, and Teruo Ono

*Institute for Chemical Research, Kyoto University, Uji, 611-0011, Kyoto, Japan*

Yoshinobu Nakatani

*University of Electro-communications, Chofu, 182-8585, Tokyo, Japan*

**Abstract**

Transmission probability of a domain wall through a magnetic nanowire is investigated as a function of the external magnetic field. Very intriguing phenomenon is found that the transmission probability shows a significant drop after exceeding the threshold driving field, which contradicts our intuition that a domain wall is more mobile in the higher magnetic field. The micromagnetics simulation reveals that the domain wall motion in the wire with finite roughness causes the dynamical pinning due to the Walker breakdown, which semi-quantitatively explains our experimental results.

The fabrication of magnetic nanowires allows us to address a single domain wall dynamics as various experimental methods are being applied for it such as giant magnetoresistance (GMR) [1-3], anisotropic magnetoresistance (AMR) [4] and tunneling magnetoresistance (TMR) [5] effects in cooperation with magnetic force microscopy [6], transmission microscopy [7] and magneto-optical Kerr effect [8,9]. It has been demonstrated that domain walls can be moved not only by a magnetic field but also by an electric current [4, 6-8, 10-24]. Since novel applications based on the domain wall motion have been proposed [25-28], the study of dynamics of a magnetic domain wall in a magnetic nanowire is of importance not only from the viewpoints of exciting fundamental physics but also from technological points of view. For the realization of such spintronic devices, the domain wall pinning by finite roughness, which inevitably exists in actual devices, is a central issue to overcome.

In this Letter, we show that the unusual dependence on the magnetic field is observed in the process of pinning and depinning of a domain wall. We found that the transmission probability of a domain wall through the magnetic nanowire displays a significant drop after exceeding the threshold driving field value. This peculiar behavior contradicting the naïve picture that a domain wall should be more mobile in the higher magnetic field is discussed based on the results of the micromagnetics simulation taking into account the edge roughness and the thermal activation. It is found that the switching of the core magnetization in a vortex domain wall due to the Walker breakdown [29] enhances the pinning probability of the moving wall, suggesting that the dynamical pinning of the domain wall plays an important role in the operation of domain-wall-based spintronics devices.

Figure 1(a) is an SEM image of the sample with the experimental setup. The samples are Permalloy ($Ni_{81}Fe_{19}$) nanowires fabricated by electron beam lithography and the lift-off method. The substrate is non-doped silicon. Wires of three different widths (190, 240, and 290 nm) were prepared with the constant thickness of 33 nm. The electrical contacts attached to the wire are layered Au(80 nm)/Cr(10 nm) electrodes. The electrodes are labeled *A*, *B* and *C*, and they are the delimiters for the two areas of the wire: the straight one (*A-B*) and the bent one (*B-C*). The length of the straight area of each wire is 23 µm.

Before each measurement the wire is reset to a single-domain state by applying a magnetic field of -520 Oe in the *H* direction (Fig. 1(b)) to ensure that there is no domain wall present in the wire. Then a small constant driving magnetic field is set in the counter direction. After that, a pulse current with the duration of 100 ns is applied though the electrode *A* to nucleate a domain wall by the Oersted field, value of which is estimated to be 400 Oe directly under the electrode *A* (Fig. 1(c)). A nucleated domain wall propagates along the wire and stops at the bent area (*B-C*) when a driving field is large enough, while it remains in the straight area (*A-B*) for a small driving field. It is easily detected whether a domain wall is in the area (*A-B*) or in the area (*B-C*) by measuring the resistances of these areas, since the resistance is decreased by the existence of a domain wall due to the anisotropic magnetoresistance effect (Figs.1(d) and 1(e)) [4]. Thus, we measure the resistances of the two wire areas before and after the electric current pulse is applied. A small DC current of 100 µA is used for measuring the resistance of the wire, so that its influence on the domain wall is negligible. The measurement is repeated 100 times for each driving field value.

Figure 2(a) shows the resistance changes for the 190 nm-width wire in the successive 100 times experiments under the driving fields of 1.1, 13.2, and 20.8 Oe. $\Delta R_1$ indicates the resistance change of the area (*A-B*), while $\Delta R_2$ is for the area (*B-C*). It should be noticed that the resistance change is almost constant, which ensures that not only a single domain wall is introduced but also the structure of the domain wall is almost the same in every experiment. For the driving field of 1.1 Oe, $\Delta R_1$ is negative and $\Delta R_2$ is zero for all experiments, indicating that a nucleated domain wall is pinned in the area (*A-B*) in every experiment. On the other hand, for the driving field of 13.2

Oe, $\Delta R_1$ shows zero and $\Delta R_2$ is negative except for the experiments #63 and #82, suggesting that a nucleated domain wall propagates through the area (*A-B*) with the transmission probability of 98 %.

Figure 2 (b) shows the transmission probabilities as a function of a driving field for the wires with different widths. Every wire examined in this study shows the following feature. The transmission probability increases abruptly for a value of the driving field close to 10 Oe and reaches 100 %. Surprisingly, further increase of the magnetic field results in the reduction of the transmission probability (going even below 50 % in the case of the narrowest wire). After taking its minimum, the probability starts to rise again, and eventually reaching 100 % for high magnetic field values (~30 Oe and more). The magnetic field for the first abrupt increase of the transmission probability can be attributed to the depinning field of a domain wall in the wire above which a domain wall can escape from the pinning potentials distributed in the wire. Puzzling is why the transmission probability decreases with the further increase of the driving field. The finding of this peculiar phenomenon against our intuition that the larger magnetic field can drive a domain wall more easily is the central experimental result in this Letter.

To elucidate this paradoxical behavior, micromagnetics simulations have been performed. The cross-sectional dimension of the wire in the simulation is the same as the narrowest wire in the experiments: the averaged wire width and the thickness are 190 nm and 33 nm, respectively. The length of the wire is 40 μm. The edge roughness of the wire has been taken into account in the simulation to realize the domain wall pinning. To produce a natural roughness, we described the roughness by Voronoi cell method with 2nm of the averaged grain size, <D> [30]. Thermal noise of 300 K has been also taken into account to reproduce the stochastic pinning and depinning of a domain wall [31]. The simulation has been repeated 16 times for each value of the magnetic field with the different sequences of the thermal noise. Each simulation was performed until the domain wall completely propagated through the wire or stopped in the wire for more than 100 ns. The two-dimensional calculations were performed by dividing the wire into rectangular prisms of $4 \times 4 \times 33$ nm$^3$; the magnetization was assumed to be constant in each prism. The typical material parameters for Permalloy were used: saturation magnetization $M_s = 1$ T, exchange stiffness constant $A = 1.0 \times 10^{-11}$ J/m, and damping constant $\alpha = 0.01$.

Figure 3 shows the simulated transmission probability as a function of a driving field. The simulated transmission probability reproduces the overall feature observed in the experiments; the abrupt increase just above the depinning field, the significant drop after showing the maximum, and the almost linear increase in the high magnetic field region. The small difference between the experiment and the simulation in the pinning field and the peak position of the transmission probability are most likely to be caused by the difference in the wire edge roughness.

The most intriguing finding in this study, the significant drop of the transmission probability with increasing a driving field, can be explained by the peculiar domain wall dynamics revealed by the simulation in the following way. We focus on a trajectory of a vortex core, since a domain wall in the wire investigated has a vortex structure. The black line in Fig. 4 indicates the calculated trajectory of a vortex core of the moving domain wall in the wire for a driving field of 18 Oe where the transmission probability in Fig. 3 takes the maximum. In this case, the domain wall moves steadily without any drastic change of domain wall structure, keeping the distance between the core and the edge almost constant. The fluctuation of the domain wall position in the y-axis (width direction of the wire) results from the effects of the edge roughness and the thermal noise.

On the other hand, in the case of a driving field of 24 Oe where the transmission probability takes the minimum, the domain wall propagation through the wire is accompanied by the alternating transverse motion of the core between both edges of the wire as shown by the green and the blue lines in Fig. 4. Here, the green and the blue lines correspond to the trajectories of the core with upward and downward magnetization, respectively. The core with upward magnetization shows the opposite transverse motion to the core with downward magnetization. Thus, the domain wall propagates though the wire with successive switching of the core magnetization, which corresponds to the Walker breakdown for the vortex domain wall [32]. Just after the switching of the core magnetization, the domain wall velocity is reduced and the core moves transverse direction of the wire. If there is a relatively large roughness, the domain wall is pinned during this slow transverse motion. This happens in this particular simulation at $x_c$ = 7.5 μm after the fifth switching of the core magnetization. This kind of dynamical pinning occurred 13 times in 16 simulations for a driving field of 24 Oe, and the pinning positions were distributed over the 40 μm-long wire.

We have shown that the Walker breakdown plays an important role in the dynamical pinning process of the moving domain wall in a magnetic nanowire. The switching of the core magnetization in a vortex domain wall due to the Walker breakdown enhances the pinning probability of the moving domain wall, resulting in the peculiar phenomena that the transmission probability of the domain wall through the wire decreases with increasing an external magnetic field.

The present work was partly supported by NEDO Spintronics nonvolatile devices project. HT acknowledges support from JSPS Research Fellowship for Young Scientists.

**Figure captions**

Figure 1

(a) SEM image of the sample with the experimental setup. (b) Initialization. (c) Domain wall introduction. (d) and (e) Relationship between the position of the domain wall and the sign of $\Delta R$.

Figure 2

(a) Resistance changes for the 190 nm-width wire in the successive 100 times experiments under the driving fields of 1.1, 13.2, and 20.8 Oe.

(b) Transmission probabilities as a function of a driving field for the wires with different widths.

Figure 3

Transmission probability as a function of a driving field calculated by the micromagnetics simulation.

Figure 4

Trajectories of a vortex core of a moving domain wall in the wire calculated by the micromagnetics simulation. The black line indicates the calculated trajectory for a driving field of 18 Oe where the transmission probability in Fig. 3 shows the peak. The green and the blue lines correspond to the trajectories of the core with upward and downward magnetization, respectively, in the case of a driving field of 24 Oe where the transmission probability takes the minimum.

(a)

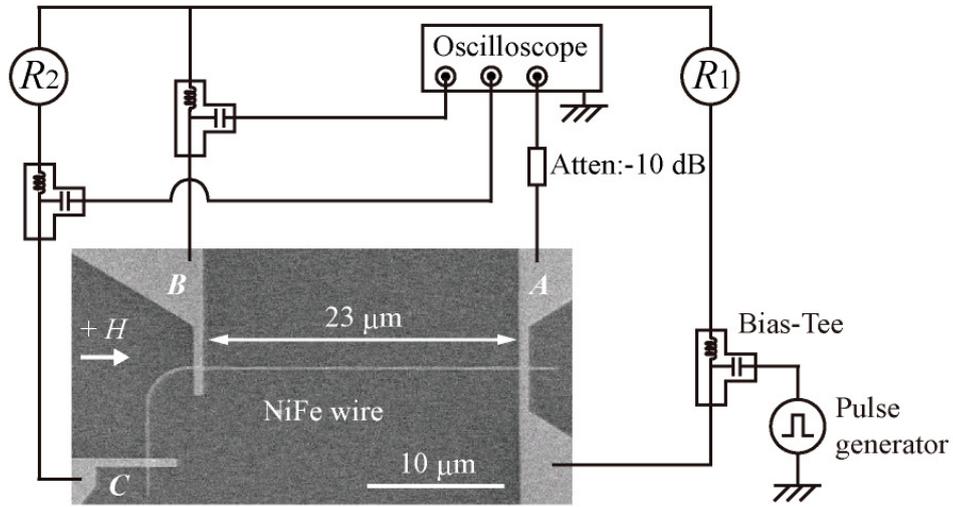

(b)

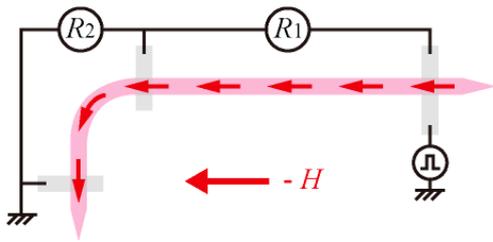

(c)

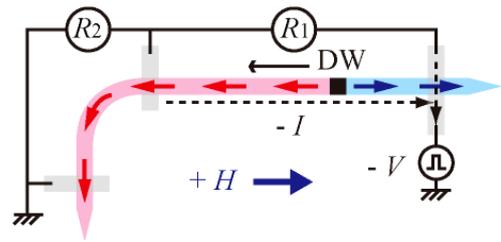

(d)

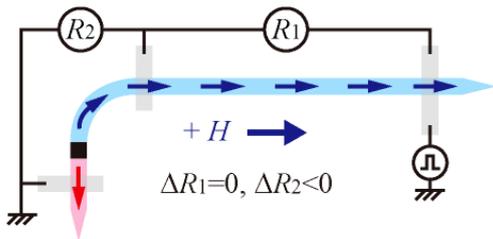

(e)

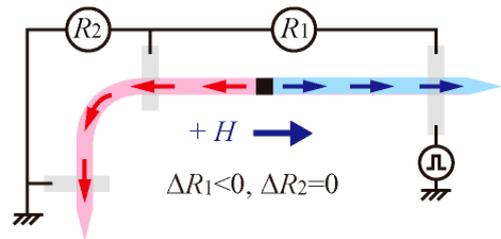

Fig. 1 H. Tanigawa

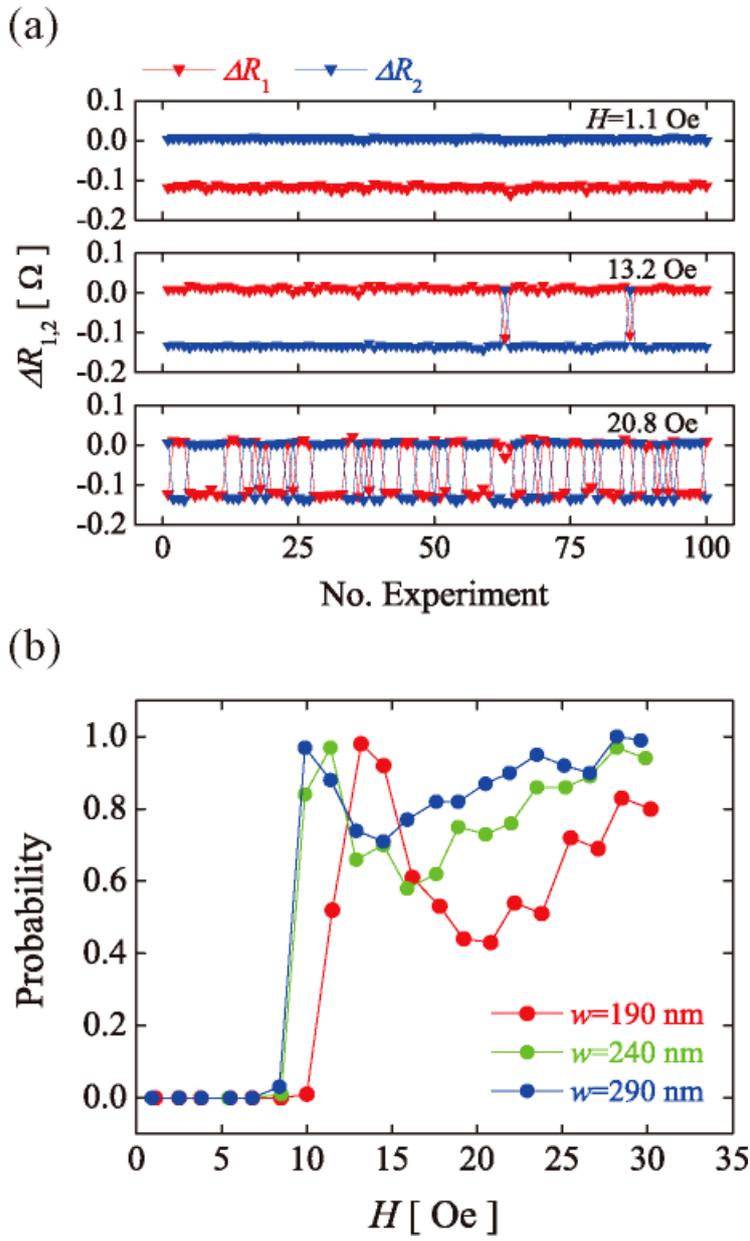

Fig. 2 H. Tanigawa

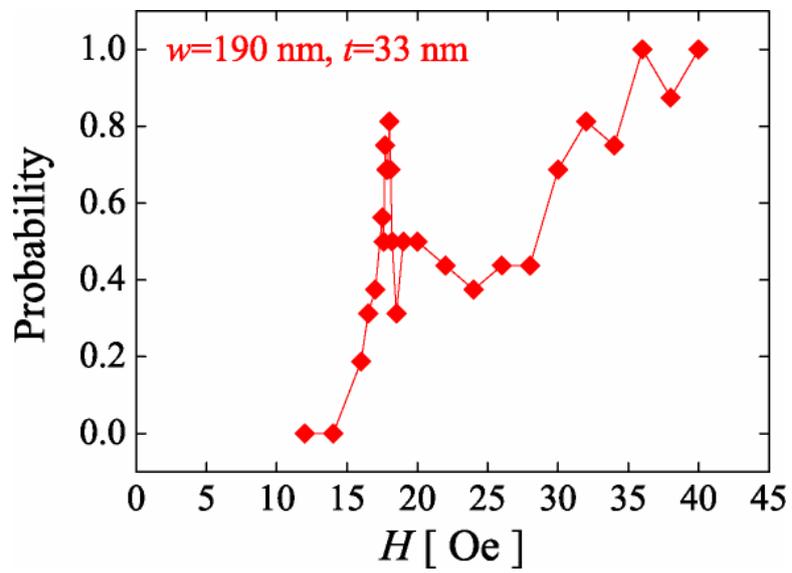

Fig. 3 H. Tanigawa

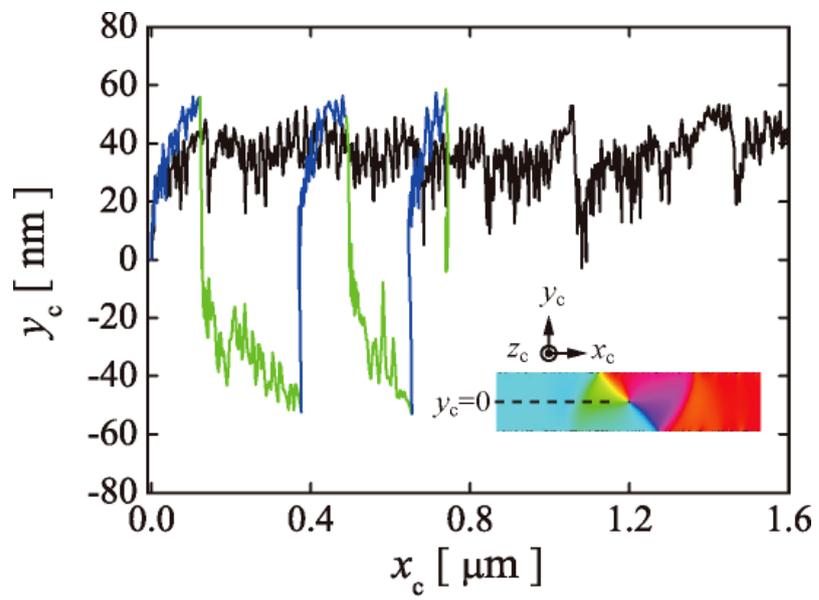

Fig. 4 H. Tanigawa